\documentclass[aps,nofootinbib,11pt]{revtex4-1}

\usepackage{epsfig}
\usepackage{amsmath,amssymb,amsfonts}
\usepackage{color}
\usepackage{hyperref}

\newcommand{\be}{\begin{equation}}
\newcommand{\en}{\end{equation}}
\newcommand{\bea}{\begin{eqnarray}}
\newcommand{\ena}{\end{eqnarray}}

\newcommand{\hbo}{\hbox to 1 true cm {\hfill } }

\newcommand{\ee}{\end{equation}}
\newcommand{\eea}{\end{eqnarray}}
\newcommand{\bml}{\begin{mathletters} \baselineskip 10pt}
\newcommand{\eml}{\baselineskip 12pt \end{mathletters}}

\newcommand{\bra}{\langle}
\newcommand{\ket}{\rangle}
\newcommand{\longl}{{\scriptscriptstyle \parallel}}

\def\lambdabar{\protect\@lambdabar}
\def\@lambdabar{%
\relax \bgroup
\def\@tempa{\hbox{\raise.73\ht0
\hbox to0pt{\kern.2\wd0\vrule width.7\wd0
height.1pt depth.1pt\hss}\box0}}%
\mathchoice{\setbox0\hbox{$\displaystyle\lambda$}\@tempa}%
{\setbox0\hbox{$\textstyle\lambda$}\@tempa}%
{\setbox0\hbox{$\scriptstyle\lambda$}\@tempa}%
{\setbox0\hbox{$\scriptscriptstyle\lambda$}\@tempa}%
\egroup }

\newcommand{\vcb}[1]{\mbox{\bf #1}}

\newcommand{\vc}[1]{\mbox{\boldmath$#1$}}

\begin{document}

\title{Covariant Worldline Numerics for Charge Motion with Radiation Reaction}

\author{Chris Harvey}
\affiliation{Department of Physics, Ume\aa\ University, SE-901 87 Ume\aa,
  Sweden}

\author{Thomas Heinzl}

\author{Nicola Iji}

\author{Kurt Langfeld}
\affiliation{School of Computing and Mathematics, University of Plymouth,
Plymouth PL4 8AA, UK}


\begin{abstract}
We develop a numerical formulation to calculate the classical motion of
charges in strong electromagnetic fields, such as those occurring in
high-intensity laser beams.  By reformulating the dynamics in terms of
$SL(2,\mathbb{C})$ matrices representing the Lorentz group,  our formulation
maintains explicit covariance, in particular the mass-shell condition.
Considering an electromagnetic plane wave field where the analytic solution is
known as a test case, we demonstrate the effectiveness of the method for
solving both the Lorentz force and the Landau-Lifshitz equations. The latter,
a second order reduction of the Lorentz-Abraham-Dirac equation, describes
radiation reaction without the usual pathologies.
\end{abstract}

\maketitle

\section{Introduction}

The issue of radiation reaction by now has a history spanning more than a
century. Building upon the pioneering work of Lorentz \cite{Lorentz:1909} and
Abraham \cite{Abraham:1905} the equation of motion for an accelerated charge
subject to an external field which is, at the same time, changed by the
backreaction of the emitted \textit{bremsstrahlung}, has been cast into its
final covariant form by Dirac \cite{Dirac:1938}. The result is a
\textit{third}-order equation for the particle trajectory, $x = x(\tau)$, and
is now aptly called the Lorentz-Abraham-Dirac (LAD) equation. It is impossible
to give a comprehensive list of references discussing this equation, so
suffice it to refer to the contemporary texts
\cite{Spohn:2004ik,Rohrlich:2007}.

There has been a renewed interest in this problem due to (at least) two recent
developments. First, it has been shown \cite{Spohn:1999uf} that the unphysical
features of the LAD equation, such as pre-acceleration and the existence of
runaway solutions, are absent if one eliminates the triple derivative term by
iteration resulting in the Landau-Lifshitz (LL) equation
\cite{Landau:1987}. Second, progress in light amplification technology
\cite{Strickland:1985} has led to new laser systems working at ultra-high
intensities above $10^{22}$ W/cm$^2$ implying field strengths in excess of
$10^{14}$ V/m (or $10^5$ T). Accelerating charges in fields of such extreme
magnitudes suggests that radiative reaction (which normally is a tiny effect)
may become physically relevant and, hence, experimentally observable
\cite{Koga:2005}.

Any discussion of backreaction and its consequences in a high-intensity
context has to accommodate the following physics. First, one expects electrons
in ultra-intense beams to become relativistic. Second, depending on the
experimental setting (particle acceleration, scattering etc.) one has
different frames of reference to consider such as the electron rest frame or
various lab frames. Hence, for the benefit of conceptual and practical
simplicity, it seems mandatory to develop an explicitly covariant formalism
based on four-vectors. In this case all equations will be valid in any frame,
and it will be straightforward to specialise to any of those at any point of
the calculation. Valuable discussions of this type have appeared recently
\cite{DiPiazza:2008lm,DiPiazza:2009pk,Sokolov:2010jx,Hadad:2010mt}. In
addition, to achieve a realistic and accurate picture of the physics one would
like to have (i) a powerful numerical formalism that (ii) respects covariance,
and, ideally, \textit{exactly} so. It turns out that such a formalism does
indeed exist as will be shown in detail below.

Before going into \textit{medias res} let us briefly discuss the physics in terms of a few relevant
parameters. Laser intensity is traditionally measured in terms of the
dimensionless amplitude\footnote{We use Heaviside-Lorentz units where $E$ and
  $B$ have the same physical dimensions and the Coulomb potential between
  electrons is $\hbar c \alpha/r$ with fine structure constant $\alpha =
  e^2/4\pi \hbar c = 1/137$.}
\be \label{eq:A0.INTRO}
  a_0 =  \frac{eE\lambdabar}{mc^2}  \; ,
\ee
which is the energy gain of a probe electron (charge $e$, mass $m$) upon
traversing a laser wavelength $\lambdabar = c/\omega$, in units of the
electron rest energy, evaluated in the lab frame where the r.m.s.\ electric
field and laser frequency are measured to be $E$ and $\omega$, respectively. A
manifestly Lorentz and gauge invariant definition will be given further below
(see also \cite{Heinzl:2008rh}). Note that $a_0$ is a purely classical
parameter as it does not contain $\hbar$. When $a_0$ exceeds unity the rapid
quiver motion of the electron in the laser beam becomes relativistic. In the
near future one expects to achieve $a_0$ values of the order of $10^3$,
corresponding to the ultra-relativistic regime
\cite{Vulcan10PW:2009,ELI:2009}.

To estimate the radiation loss we use Larmor's formula which expresses the
radiated power $P$ in terms of the acceleration $a = eE/m$
\cite{Jackson:1999},
\be
  P  = \frac{2}{3} \, \frac{e^2}{4\pi c^3} \, a^2 = \frac{2}{3} \hbar \alpha \, \omega^2 \, a_0^2 \; .
\ee
Thus, the energy radiated per time is proportional to $a_0^2$ and can be made
dimensionless upon dividing by $\omega mc^2$. Introducing the dimensionless
energy variable
\be
  \nu \equiv \frac{\hbar\omega}{mc^2} \; ,
\ee
we obtain the energy loss per laser cycle in units of $mc^2$,
\be \label{eq:R}
  R \equiv \frac{P}{\omega mc^2} = \frac{2}{3} \alpha \, \nu \, a_0^2 \; ,
\ee
which is precisely the parameter used in
\cite{Koga:2005,DiPiazza:2008lm,DiPiazza:2009pk}. The authors of
\cite{DiPiazza:2009pk} state that one enters the ``radiation dominated
regime'', where radiation damping can no longer be neglected, when $R$ exceeds
unity.  According to (\ref{eq:R}) this amounts to an energy loss larger than
$mc^2$ per laser cycle. However, one should also take into account the energy
gain per cycle as measured by $a_0$, cf.\ (\ref{eq:A0.INTRO}). To this end we
define the energy balance parameter
\be \label{eq:KAPPA}
  \kappa \equiv R/a_0 = \frac{2}{3} \alpha \, \nu \, a_0 \; .
\ee
which is just the ratio of energy loss and gain. Thus, when $\kappa$ becomes
of order unity the radiation loss equals the typical kinetic energy of the
accelerated charge. This has been stated before in \cite{McDonald:1998}.

This paper is organised as follows. In Section~\ref{sec:CHARGE.FIELD} we
recall the LAD equation and its reduction to the LL equation. We review the
analytic solution of the latter before we introduce our new numerical scheme
in Section~\ref{sec:num}. In Section~\ref{sec:APP} we then utilise it for solving the equation of
motion of a charge in a pulsed plane wave, both without and with radiative
reaction. The analytic solutions are reproduced to a high accuracy. Conclusions are finally presented in Section~\ref{sec:CONCL}.

\section{Coupled Charge-Field Dynamics }
\label{sec:CHARGE.FIELD}

The kinematics of a relativistic particle are encoded in its trajectory (or
worldline) $x^\mu (\tau)$, its four-velocity $u^\mu (\tau)$ and its four
acceleration $a^\mu (\tau)$ (and possibly higher derivatives
\cite{Russo:2009yd}). All these are conveniently parameterised by proper time
$\tau$. The trajectory is found by solving the equations of motion for $x^\mu
(\tau)$. As is well known, the resulting LAD equation provides an
example of a higher-derivative theory including a third derivative (or
`jerk'') term \cite{Russo:2009yd}.

\subsection{Equations of Motion}
\label{subsec:EOM}

The LAD equation follows from the usual action principle, the action being
obtained by minimally coupling a relativistic point particle to the classical
Maxwell field \cite{Jackson:1999,Landau:1987},
\be \label{eq:ACTION1}
  S = -mc^2 \int d\tau - \frac{e}{c} \int dx^\mu A_\mu - \frac{1}{4} \int d^4
  x \, F_{\mu\nu}F^{\mu\nu} \; , \quad dx^\mu = u^\mu \, d\tau \; .
\ee
Naturally, the action depends on Lorentz scalars\footnote{Our metric is
  $g_{\mu\nu} = \mbox{diag}(1,-1,-1,-1)$ implying a Minkowski scalar product
  $a \cdot b = g_{\mu\nu} a^\mu b^\nu = a^0 b^0 - \vcb{a} \cdot \vcb{b}$.}
only and hence is relativistically invariant. Note that particle and field
aspects are manifest in the different integration measures, $d\tau$ and $d^4
x$, respectively. In this context, the second term describing the interaction
of charged particle and field is somewhat of a ``hybrid''. It can be rewritten
with a ``field theoretic'' measure as follows. A particle moving along the
trajectory $x^\mu (\tau)$ amounts to a four-current
\be \label{eq:CURRENT}
  j^\mu (x) = \int d\tau \, u^\mu (\tau) \, \delta^4 (x - y(\tau)) \; .
\ee
Plugging this into (\ref{eq:ACTION1}) we obtain the alternative representation
\be \label{eq:ACTION2}
  S = -mc^2 \int d\tau - \frac{e}{c} \int d^4 x \, j^\mu A_\mu - \frac{1}{4}
  \int d^4 \, x F_{\mu\nu} F^{\mu\nu} \; .
\ee
Written either way, the action is invariant under gauge transformations,
$A_\mu \to A_\mu + \partial_\mu \chi$ with an arbitrary scalar function
$\chi$.

Varying the action with respect to $A_\mu$ and $x_\mu$ yields a coupled system
consisting of Maxwell's equations,
\be \label{eq:MAXWELL}
  \partial _\mu F^{\mu \nu }  = j^\nu  \; ,
\en
and the relativistic generalisation of Newton's second law,
\be \label{eq:LORENTZ2}
  m \; \dot{u}_\mu  =  \frac{e}{c} \, F_{\mu \nu } \, u^\nu \equiv F_\mu \; ,
\en
where the Lorentz four-force $F_\mu$ appears on the right-hand side. Following
Dirac \cite{Dirac:1938} one eliminates $F_{\mu\nu}$ from (\ref{eq:LORENTZ2})
according to
\be
  F^{\mu\nu} = F^{\mu\nu}_{\mathrm{in}} + F^{\mu\nu}_{\mathrm{rad}} \; ,
\ee
where the homogeneous and inhomogeneous solutions, $F^{\mu\nu}_{\mathrm{in}}$
and $F^{\mu\nu}_{\mathrm{rad}}$, of the wave equation (\ref{eq:MAXWELL})
represent the prescribed external field (or ``in-field'') and the radiation
field, respectively. The calculation of $F^{\mu\nu}_{\mathrm{rad}}$ is
somewhat tedious. As the radiation field diverges on the particle world-line
one encounters a  short-distance singularity which is removed by mass
renormalisation (albeit in a classical context) \cite{Dirac:1938}. A nice
exposition may be found in Coleman's paper \cite{Coleman:1982} (in particular
Sect.~6). For the sake of simplicity we will henceforth use the same letter
$m$ to denote the renormalised (observable) electron mass.

The final upshot is the celebrated LAD equation,
\be \label{eq:LAD1}
  m \dot{u}^\mu =  \frac{e}{c} F_{\mathrm{in}}^{\mu\nu} u_\nu  -  \frac{2}{3}
  \frac{e^2}{4\pi c^5} (u^\mu \ddot{u}^\nu - u^\nu \ddot{u}^\mu) \,
  u_\nu \; ,
\en
in the form first presented by Dirac \cite{Dirac:1938}. Note that the tensor
multiplying $u_\nu$ is manifestly antisymmetric in $\mu$ and $\nu$. Thus, $u
\cdot \dot{u}$ remains zero as is required by the space-like nature of
acceleration, $\dot{u}^2 < 0$. Taking the second proper-time derivative of
$u^2 = c^2$,
\be
  \frac{1}{2} \frac{d^2 u^2}{d\tau^2} = \ddot{u} \cdot u + \dot{u} \cdot
  \dot{u}  = 0 \; ,
\ee
we may equivalently write (\ref{eq:LAD1}) (henceforth omitting the subscript
``in'') as
\be \label{eq:LAD2}
  m \dot{u}^\mu = \frac{e}{c} F^{\mu\nu} u_\nu + \frac{2}{3} \frac{e^2}{4\pi
    c^3} (\ddot{u}^\mu + \dot{u}^2 \, u^\mu/ c^2) \; ,
\en
where the antisymmetry on the right-hand side is no longer manifest. The
appearance of the notorious $\ddot{u}$ term in (\ref{eq:LAD1}) and
(\ref{eq:LAD2}) leads to pathologies such as runaway solutions and/or
pre-acceleration which have been discussed in the literature for decades (see
e.g.\ the texts \cite{Rohrlich:2007,Spohn:2004ik}). An elegant (and
consistent!) way to remove the unwanted features is to replace $\ddot{u}^\mu$
and $\dot{u}^2$ with the help of the leading (i.e., Lorentz) term in the
equation of motion (\ref{eq:LAD1}) thus ``reducing the order''
\cite{Gralla:2009md} to obtain the Landau-Lifshitz (LL) equation
\cite{Landau:1987},
\be \label{eq:LL}
  m \dot{u}^\mu = \frac{e}{c} F^{\mu\nu} u_\nu + \frac{2}{3} \frac{e^2}{4\pi
    c^3} \left\{
  \frac{e}{mc} \dot{F}^{\mu\nu} u_\nu + \frac{e^2}{m^2 c^2}
  F^{\mu\alpha}F_\alpha^{\;\;\nu} u_\nu - \frac{e^2}{m^2 c^4} u_\alpha
  F^{\alpha\nu}F_\nu^{\;\;\beta} u_\beta \, u^\mu \right\} \; .
\en
This equation is much ``better behaved'' than (\ref{eq:LAD1}) and
(\ref{eq:LAD2}) as the right-hand side only involves velocities and no higher
derivatives. It has recently been rederived using rigorous geometric
perturbation theory (or classical renormalisation group flow) in
\cite{Spohn:1999uf} and by a sophisticated limiting procedure (which also
produces electric and magnetic moment contributions) in
\cite{Gralla:2009md}. A promising analysis using the language of effective field theory has been presented in \cite{Galley:2010es}

For what follows it is useful to adopt a notation in terms of dimensionless
variables choosing appropriate units. These will also come in handy for
the numerical approach of the next section.

\subsection{Dimensionless variables}
\label{subsec:dimless}

We assume that our laser beam is described by a light-like wave vector $k =
(\omega/c, \vcb{k})$, $k^2 = \omega^2/c^2 - \vcb{k}^2 = 0$, with $\omega$ and
$\vcb{k}$ being lab frame coordinates. To combine this with the electron
motion we follow Wald \cite{Wald:1984rg} and define a frequency by dotting $k$
into the initial velocity, $u_0$,
\be \label{eq:OMEGA0}
  \Omega_0 \equiv k \cdot u_0 \; .
\ee
The rationale here is the that in an experiment one expects to be in control
of initial conditions such as the initial velocity, $u_0$. If the particle is
initially at rest (defining the ``initial rest frame'' or IRF) we have
$u_0^\mu = c \delta^\mu_{\;\;0}$ and $\Omega_0 = \omega_0$, with $\omega_0$
denoting the laser frequency in the IRF.

We also define a dimensionless proper time, $s \equiv \Omega_0 \tau$, and
adopt natural units, $ \hbar = c = 1$, unless otherwise stated. To avoid
clumsy notation, we will henceforth denote $s$-derivatives by an
over-dot. Finally, we rescale $e F^{\mu\nu}/me\Omega_0 \to F^{\mu\nu}$ which also renders $F^{\mu\nu}$ dimensionless. In this new notation the parameters of the
introduction may be given invariant definitions according to
\bea
  a_0^2 &=& u_{0\mu} \bra F^{\mu\alpha}F_{\alpha}^{\phantom{\alpha}\nu} \ket
  u_{0\nu} \; ,  \\
  \nu_0 &\equiv& \frac{\Omega_0}{m} \;  .
\eea
The brackets $\bra \ldots \ket$ at this point denote a \emph{typical} value
such as the root-mean-square (proper time average) or the amplitude (cycle
maximum). This implies that $F^{\mu\nu}$ is proportional to
$a_0$ which will be made explicit later on. It is useful to introduce a
(dimensionless) energy density
\be
  w(u) \equiv u_{\mu}  F^{\mu\alpha}F_{\alpha}^{\phantom{\alpha}\nu} u_{\nu} \; ,
\ee
and $w_0 \equiv w(u_0)$ such that $a_0^2 = \bra w_0 \ket$. Finally, we define
the effective coupling parameter
\be
  r_0 \equiv \frac{2}{3} \alpha \nu_0 \; ,
\ee
which will appear repeatedly, such that $\kappa = r_0 a_0$ and $R = r_0
a_0^2$, cf.\ (\ref{eq:R}) and (\ref{eq:KAPPA}).

With these prerequisites, the LAD and LL equations may be compactly written as
\bea
  \dot{u}^\mu &=& F^{\mu\nu} u_\nu + r_0 (\ddot{u}^\mu + \dot{u}^2 u^\mu) \;
  , \label{eq:LAD.DIMLESS1}\\
  \dot{u}^\mu &=& F^{\mu\nu} u_\nu + r_0 (\dot{F}^{\mu\nu} +
  F^{\mu\alpha}F_{\alpha}^{\phantom{\alpha}\nu} - w \, g^{\mu\nu}) u_\nu
 \; . \label{eq:LL.DIMLESS1}
\eea
For our numerical approach it will be important to have manifestly
antisymmetric tensors on the right-hand side, so we provide these forms as
well,
\bea
  \dot{u}^\mu &=& \Big\{ F^{\mu\nu} + r_0 ( \ddot{u}^\mu u^\nu - \ddot{u}^\nu
  u^\mu ) \Big\} u_\nu  \; , \label{eq:LAD.DIMLESS2} \\
  \dot{u}^\mu &=& \left\{ F^{\mu\nu} + r_0 \left[ (\dot{F}^{\mu\beta} +
    F^{\mu\alpha}F_{\alpha}^{\phantom{\alpha}\beta}) u_\beta \, u^\nu - (\mu
    \leftrightarrow \nu) \right] \right\} u_\nu \label{eq:LL.DIMLESS2}
 \; .
\eea
It is crucial to note that the LL equation is an expansion in powers of $r_0$
(or $\alpha$), with coefficients being proportional to powers of field
strength, hence $a_0$. Obviously, the leading order ($r_0^0$) is the Lorentz
term while the LL term is $O(r_0)$.

\subsection{Analytic Solution for a Pulsed Plane Wave}
\label{subsec:planewave}

Things simplify further upon modelling the laser beam by a plane wave. In this
case the field strength solely depends on the invariant phase, $F^{\mu\nu} =
F^{\mu\nu} (k \cdot x) \equiv F^{\mu\nu} (\phi)$, and is assumed to be transverse,
\be \label{eq:TRANSVERSE}
  k_\mu F^{\mu\nu} = 0.
\ee
Most importantly, plane wave fields are \textit{null fields}
\cite{Synge:1935zzb,Stephani:2004ud} which are characterised by peculiar
Lorentz properties. The standard field invariants vanish,
\be
  F_{\mu\nu} F^{\mu\nu} = F_{\mu\nu} \tilde{F}^{\mu\nu} = 0 \; ,
\ee
where $\tilde{F}^{\mu\nu}$ denotes the dual field strength. This implies that
the energy momentum tensor is just the (matrix) square of $F^{\mu\nu}$,
\be \label{eq:TMUNU}
  T^{\mu\nu} = F^{\mu\alpha}F_\alpha^{\;\;\nu} \; ,
\ee
implying, for instance, that $w = u_\mu T^{\mu\nu} u_\nu$ which shows that $w$
is indeed the energy density of the wave as measured in the instantaneous
electron rest frame. In addition, we see that the LL equation in the form
(\ref{eq:LL.DIMLESS1}) or (\ref{eq:LL.DIMLESS2}) depends explicitly on
$T^{\mu\nu}$. The matrix cube of $F^{\mu\nu}$, and hence all higher powers,
vanish. This will become important later on. It turns out that
for plane wave a solution for the particle trajectory can be found
in terms of a few integrals (which are then evaluated numerically).

All these features make the case of a plane wave an ideal testing ground
for our new numerical method which is detailed in the next section.
To be specific, we introduce the null vector $n^\mu = k^\mu/\Omega_0$
(implying $n \cdot u_0 = 1$) and two space-like
polarisation vectors, $\varepsilon_i^\mu$, with the scalar products
\be \label{eq:ortho}
n^2 = 0 \; , \quad n \cdot \varepsilon_i = 0 \; , \quad \varepsilon_i \cdot
\varepsilon_j = - \delta_{ij}  \; .
\en
It is convenient to write the field strength in  terms of profile functions
$f_i (\phi)$ and  elementary tensors, $f_i^{\mu\nu}$, multiplying the strength
parameter $a_0$,
\be \label{eq:F.MU.NU}
  F^{\mu\nu} (\phi ) = a_0 \, f_i (\phi) \, f_i^{\mu\nu} \, , \hbo
  f_i^{\mu\nu} = n^\mu \varepsilon_i^\nu - n^\nu \varepsilon_i^\mu .
\ee
For simplicity, we immediately specialise to linear polarisation,
\be
f_2 =0 , \hbo f_1 \equiv f, \hbo n^\mu = (1, \hat{\vcb{z}}), \hbo
\varepsilon_1^\mu = (0, \hat{\vcb{x}}) \; ,
\label{eq:PULSE_dir}
\en
with the choice of $n^\mu$ corresponding to a particle initially at rest
($\Omega_0 = \omega_0$). In addition, we choose a pulse with a Gaussian
envelope,
\be
  f(\phi) \equiv - \exp \left\{ - \frac{(\phi - \phi_0)^2}{N^2} \right\} \,
\sin (\phi) \; ,
\label{eq:PULSE}
\en
where, obviously, $\phi_0$ denotes the centre of the pulse while $N$ counts the
number of cycles within.

It turns out that the null-plane properties of plane waves are sufficiently
strong to still allow for an analytic solution
\cite{DiPiazza:2008lm,Hadad:2010mt}, the basic technical reason being the
decoupling of the light-cone component $u_\longl \equiv n \cdot u$ in the LL
equation. We briefly recapitulate the main steps in our condensed notation
(\ref{eq:F.MU.NU}), which we plug into (\ref{eq:LL.DIMLESS1}) to obtain
\bea
  \dot{u}^\mu &=&  \left[ a_0 f \, f^\mu _{\phantom{\mu} \nu} + r_0 a_0 \, u_\longl \,
  \left\{ f' \, f^\mu _{\phantom{\mu} \nu} + a_0 f^2 (n^\mu u_\nu - n_\nu  u^\mu )
  \right\} \right] \; u^\nu \; ,
  \label{eq:k1}
\ena
the prime denoting derivatives with respect to the invariant phase,
$\phi$. Dotting in $n$, and using transversality, $n_\mu f^\mu _{\phantom{\mu}
  \nu} =0$, see (\ref{eq:TRANSVERSE}) and (\ref{eq:ortho}), we indeed find
that the dynamics of the longitudinal (or light-cone) component $u_\longl = n
\cdot u = \dot{\phi}$ decouples,
\be
  \dot{u_\longl }  =  - \; r_0 a_0^2 \, f^2(\phi) \, u_\longl ^3 \; .
\label{eq:k2}
\en
The latter equation is easily integrated by separating of variables,
\be
  u_\longl (\phi )  =  \frac{ 1 }{ 1 + r_0 I(\phi) } , \hbo I(\phi )  \equiv  a_0^2 \, \int _0^\phi f^2(\varphi) \; d\varphi ,
\label{eq:k3a}
\en
where we have taken into account the initial conditions $u_\longl = 1$. Note
that $\dot{\phi} = u_\longl (\phi ) > 0$ implying a monotonic relation between
$\phi $ and (rescaled) proper time $s$,
\be
  s (\phi ) \; = \; \int_0^\phi d\varphi \, \Bigl[ 1 \, + \, r_0 \, I(\varphi
    ) \Bigr] \;  \; .
\label{eq:k3b}
\en
This may be used to trade proper time $s$ for invariant phase $\phi$, $\dot{h}
= h' u_\longl$ for any function $h$ of $s = s(\phi)$. Following~
\cite{DiPiazza:2008lm}, we introduce a rescaled velocity $v^\mu $ by
\be \label{eq:UV}
  u^\mu (s) = u_\longl (\phi) \, v^\mu (\phi) ,
\ee
such that the acceleration $\dot{u}$ turns into
\be
  \dot{u}^\mu  \; = \; - \, r_0 a_0^2  \, f^2(\phi) \, u_\longl ^3 \, v^\mu
  (\phi) \; + \; u_\longl ^2 (\phi) \, v^{\prime \mu}  .
\label{eq:k3}
\en
Inserting this into (\ref{eq:k1}), we find
\be
  v^{\prime \mu} = \left[ \frac{a_0 \, f(\phi ) }{ u_\longl (\phi ) } + r_0
    a_0 f^\prime (\phi ) \right] \, f^\mu _{\phantom{\mu} \nu} \, v^\nu \; +
  \; \frac{ r_0 a_0^2 \, f^2(\phi) }{ u_\longl (\phi )} \, n^\mu \; .
\label{eq:k4}
\en
Due to transversality (\ref{eq:TRANSVERSE}) the inhomogeneous longitudinal
term is in the kernel of $f^\mu _{\phantom{\mu} \nu}$ so that the solution of
(\ref{eq:k4}) is provided by the ansatz
\bea
v^\mu (\phi ) &=& \left[ \exp ( I_1(\phi) \, f) \right] ^\mu _{\phantom{\mu} \nu}
\, v^\nu_0 \; + \; I_2(\phi) \, n^\mu .
\label{eq:k5}
\eea
The ``shape integrals'',
\bea
  I_1(\phi) &=& \int _0 ^\phi d\varphi \; \left[ \frac{a_0 \, f(\varphi ) }{
  u_\longl (\varphi ) } + r_0 a_0 f^\prime (\varphi ) \right] , \hbo
  I_2(\phi ) \; = \; \int _0 ^\phi d\varphi \;
  \frac{ r_0 a_0^2 \, f^2(\varphi) }{ u_\longl (\varphi )} \; ,
\label{eq:k6}
\ena
have been chosen such that the initial condition $v(0)= v_0 = u_0$ is
maintained. The null field properties encoded in (\ref{eq:ortho}) lead to
$$
(f^2) ^\mu _{\phantom{\mu} \nu} \; = \; n^\mu \, n_\nu, \hbo
(f^n) ^\mu _{\phantom{\mu} \nu} =0 \; , \; \; n \ge 3 ,
$$
and greatly simplify the evaluation of exponential in (\ref{eq:k5}),
\be
  v^\mu (\phi ) \; = \;   v^\mu_0 \; + \; I_1(\phi) \, f ^\mu _{\phantom{\mu} \nu}
  \, v^\nu_0 \; + \; \Bigl[ I_2(\phi) + \frac{1}{2} I_1^2(\phi) \;
   \Bigr] \, n^\mu  .
\label{eq:k7}
\en
In summary, we find the compact solution for the four velocity
\be
u^\mu (s) \; = \; \frac{ v^\mu (\phi) }{ 1 + r_0 \, I(\phi) }  ,
\label{eq:k8}
\en
with $v^\mu$ given in (\ref{eq:k7}) and proper time $s = s(\phi)$ according to
(\ref{eq:k3b}).

\section{Numerical studies of the particle motion }
\label{sec:num}

\subsection{Rationale}

A typical numerical approach to solving the equation of motion
(\ref{eq:LORENTZ2}), an ordinary differential equation, would be a finite
difference scheme where proper time is discretised into intervals of length
$ds$. In this case, one generically expects a violation of the on-shell
condition, $u^2 = 1$, of power law form,
\be \label{eq:UU.VIOLATION}
  \frac{d}{ds} \, u^2 = 2 \, u \cdot \dot{u} = \mathcal{O} \left( ds^p
  \right) \ne 0 \; , \hbo p > 0 \; ,
\ee
where the power $p$ increases with the order of the scheme. This may be viewed
as yet another instance of the violation of the product (Leibniz) rule upon
discretising derivatives. At best, the induced error can be viewed as a mass
shift which, however, depends on proper time,
\be
  p^2 = m^2 u^2 \to m^2 (1 + K \, ds^p) \; ,
\ee
with some constant $K$. Obviously, this error is going to interfere with the
exact dynamics, for instance, according to (\ref{eq:UU.VIOLATION}), $u \cdot
\dot{u} \ne 0$ and the acceleration $\dot{u}$ will no longer be space-like.

To avoid such Lorentz violations we will develop a numerical scheme which is
manifestly covariant and which exactly incorporates the on-shell condition,
$u^2 = 1$.


\subsection{Matrix Dynamics}

Our numerical approach is to some extent inspired by lattice gauge theory
where dynamical variables, say $x$, with values in a Lie algebra are traded
for \emph{group} valued degrees of freedom, symbolically denoted by $X =
\exp(ix)$.

In a first step we employ the well known equivalence (modulo $\mathbb{Z}(2)$)
of the Lorentz group with the group $SL(2, \mathbb{C})$ of complex 2-by-2
matrices with unit determinant (see e.g.\ \cite{Weinberg:1995mt}, Ch.~2.7). Introducing
the matrix basis $\sigma_\mu \equiv (\mathbb{I}, \vc{\sigma})$ with
$\vc{\sigma}$ denoting the three Pauli matrices, we may associate the four
velocity $u^\mu$ with the hermitian matrix
\be \label{eq:U.SL2C}
  U \equiv u^\mu \sigma_\mu = \left( \begin{array}{cc} u^+ & v_- \\ v_+ &
    u^-   \end{array}\right) \; , \quad u^\pm = u^0 \pm u^3 \; , \quad v_\pm
  \equiv u^1 \pm iu^2 = v_\mp^*\; .
\ee
The invariant square of $u$ then becomes
\be
  u^2 = \det (U) = u^+ u^- - v_+v_- = 1 \; .
\ee
A Lorentz transformation of $u$, $u^{\mu} \to \lambda^{\mu}_{\;\;\nu} u^\nu$,
may then be implemented as an $SL(2, \mathbb{C})$ conjugation,
\be
  U \to \Lambda U \Lambda^\dagger = (\lambda^{\mu}_{\;\;\nu} u^\nu)\sigma_\mu
  \; , \quad \det(\Lambda) = 1 \; .
\ee
The condition of unimodularity guarantees the invariance of $u^2 =
\det(U)$. The significance of this for our problem is the following. For
constant fields, $F_{\mu\nu} = const$, the Lorentz equation ($r_0 = 0$) may be
written as
\be \label{eq:INF.LORENTZ}
  \dot{u}_\mu = F_{\mu\nu} u^\nu \equiv \omega_{\mu\nu} u^\nu \; , \quad
  \omega_{\mu\nu} = - \omega_{\nu\mu} \; .
\ee
But the right-hand side is just an infinitesimal Lorentz transformation
($\omega_{\mu\nu}$ being antisymmetric, and hence in the Lie algebra of the
Lorentz group). In other words the associated trajectory defines a symmetry
orbit in Minkowski space as first noted (and classified) by Taub
\cite{Taub:1948}. Hence, (\ref{eq:INF.LORENTZ}) is solved by a (finite)
Lorentz transformation,
\be \label{eq:LT.SOLUTION}
  u^\mu (s) = \lambda^\mu_{\;\;\nu} (s) \, u_0^\nu \; , \quad \lambda (s) =
  \exp(\omega s) \; ,
\ee
where $u_0^\mu$ denotes the initial four-velocity. We need to translate this
into $SL(2, \mathbb{C})$ language. For constant fields this is discussed in Ch.~1 of
\cite{Itzykson:1980rh} without maintaining explicit covariance. In what
follows we present a derivation that (i) preserves manifest covariance at any
stage and (ii) is valid for arbitrary (i.e.\ nonconstant) field strengths,
$F_{\mu\nu} = F_{\mu\nu} (x, u)$.

Our method makes use of the concept of electromagnetic duality which, among
other things, is the basic tool for an algebraic characterisation of
electromagnetic fields \cite{Stephani:2004ud} such as null fields. Recall that
the dual field strength is given by a contraction with the totally
antisymmetric Levi-Civita tensor,
\be
  \tilde{F}_{\mu\nu} = \frac{1}{2} \epsilon_{\mu\nu\rho\sigma}F^{\rho\sigma} \; .
\ee
We may therefore decompose any field strength $F_{\mu\nu}$ into its selfdual
and antiselfdual components,
\be \label{eq:AS.DECOMP}
  F_{\mu\nu} = \frac{1}{2} (F_{\mu\nu} + i \tilde{F}_{\mu\nu}) + \frac{1}{2}
  (F_{\mu\nu} - i \tilde{F}_{\mu\nu}) \equiv \Phi_{\mu\nu} + \Phi_{\mu\nu}^*
  \; ,
\ee
where self-duality in Minkowski space means \cite{Stephani:2004ud}
\be
   \tilde{\Phi}_{\mu\nu} = - i \Phi_{\mu\nu} \; .
\ee
The selfdual tensor $\Phi_{\mu\nu}$ may alternatively be written in terms of
(a Minkowski version of) `t~Hooft's symbols
\cite{'tHooft:1976fv,Schafer:1996wv,Belitsky:2000ws}
\be
  \eta_{a\mu\nu} = -i \, \epsilon_{0a\mu\nu} + g_{\mu a} \, g_{\nu 0} - g_{\mu
    0} \, g_{\nu a} \; , \quad a = 1,2,3 \; ,
\ee
which yields the compact expression
\be \label{eq:F.I}
  \Phi_{\mu\nu} \equiv F^a \eta_{a\mu\nu} \; , \quad F^a = \frac{1}{2} (E^a -
  i B^a) \; .
\ee
To proceed we insert the decomposition (\ref{eq:AS.DECOMP}) into the equation
of motion (\ref{eq:LORENTZ2}) and contract with $\sigma^\mu$. This yields
\be
  \dot{U} = F^a \eta_{a\mu\nu} \sigma^\mu u^\nu + h.c. \; ,
\ee
``$h.c.$'' denoting the hermitean conjugate. The matrix identity,
\be
  \eta_{a\mu\nu} \sigma^\mu u^\nu = \sigma_\nu \sigma_a u^\nu = U \sigma_a \; ,
\ee
together with the abbreviation  $\mathbb{E} \equiv F^a \sigma_a \in su(2)$
finally yields the desired $SL(2, \mathbb{C})$ equation of motion,
\be \label{eq:SL2C.EOM}
  \dot{U} = \mathbb{E}^\dagger U + U \mathbb{E}  \; .
\ee
The appearance of \emph{two} terms on the right-hand side reflects the fact
that Lie algebra of the Lorentz group decomposes into \emph{two} $su(2)$
subalgebras\footnote{Denoting rotation and boost generators by $L^a$ and
  $K^a$, respectively, the two algebras are generated by the linear
  combinations $L^a \pm i K^a$ which correspond to $F^a$ and its complex
  conjugate above, cf.\ (\ref{eq:F.I}).}.

Note that, in general, $\mathbb{E}$ (or $F_{\mu\nu}$) will depend on $s$,
$u^\mu (s)$ and $x^\mu (s)$. If there is only explicit $s$-dependence,
$\mathbb{E} = \mathbb{E}(s)$, then (\ref{eq:SL2C.EOM}) is similar to a
(linear) Schr{\"o}dinger equation with time-dependent Hamiltonian. Hence, it
may be solved by introducing an evolution operator, i.e., the time-ordered
product
\be
  L(s) \equiv \mathcal{T} \exp \left\{ \int_0^s ds' \, \mathbb{E}^\dagger (s')
  \right\} \in SL(2, \mathbb{C}) \; .
\ee
The solution of (\ref{eq:SL2C.EOM}) is then given by conjugation with $L$,
\be \label{eq:SL2C.SOLUTION}
  U(s) = L(s) \, U(0) \, L^\dagger (s) \; ,
\ee
which generalises the solution (\ref{eq:LT.SOLUTION}) in terms of a Lorentz
transformation.

\subsection{Numerical Implementation}
\label{subsec:SL2C}

If $\mathbb{E} = \mathbb{E}(s; x(s), u(s))$ the equation of motion (even in
the absence of backreaction) becomes nonlinear. Nevertheless, an iterative
scheme based on the above can still be expected to work. Note that the solution
(\ref{eq:SL2C.SOLUTION}) is ideally suited for the required numerical
computations. To see this, introduce a discrete set of $n+1$ equally spaced
proper time values $s_k$, $k = 0 \ldots n$, such that
\be
  s _0 =0 \; ,  \hbo s_k \, = \, k \, ds \; , \hbo   s_n = s \, , \hbo
  \mathbb{E}_k := \mathbb{E}\left( x \left( s_k \right) \right) \; .
\ee
We then approximate (with an error of order $ds^2$)
\be \label{eq:L.APPROX}
  L \approx \exp \{ \mathbb{E}^\dagger_n \, ds \} \times \ldots \times \exp \{
  \mathbb{E}^\dagger_1 \, ds \} =: L_n \; ,
\ee
where ``$\times $'' denotes matrix multiplication. For the solution
(\ref{eq:SL2C.SOLUTION}) this implies
\be
  U(s) \; = \; U_n (s) + \; {\cal O}(ds) \; , \quad \mbox{where} \quad   U_n
  (s) \, = \, L_n \, U(0) \, L_n^\dagger \; ,
  \label{eq:n10}
\en
which is the statement that our method corresponds to a first-order scheme.

It is now straightforward to verify that the on-shell condition, $u^2 = 1$, is
exactly maintained by the approximate solution. To this end, note that
(\ref{eq:L.APPROX}) represents a decomposition of the matrix $L_n$ into a
product of unimodular matrices $\exp ( \mathbb{E}^\dagger_k ds )$. So $L_n$ is unimodular as well, $\det(L_n) = 1$, hence $L_n \in SL(2, \mathbb{C})$. This finally yields
\be
  \det \, U_n (\tau) = \det \, U(0) = 1 \; .
  \label{eq:n11}
\en
and establishes that, our discretisation notwithstanding, the on-shell
condition is exactly preserved.

\section{Applications}
\label{sec:APP}

\subsection{Pulsed Plane Wave without Radiative Reaction }

The Lorentz equation for a charge in a plane wave background, $F_{\mu\nu} =
F_{\mu\nu} (k \cdot x)$, was first solved by Taub in 1948
\cite{Taub:1948}. His solution is easily rederived from our general result of
Subsection~\ref{subsec:planewave}. All we have to do is switch off radiation
reaction by setting $r_0=0$. In this case, the relationship (\ref{eq:k3b})
implies equality of (rescaled) proper time and invariant phase,
\be \label{eq:PHI.S}
  s  = \phi \; = \; k \cdot x \,  , \hbo
\ee
as well as constancy of $u_\longl$,
\be
  u_\longl \; = \; n \cdot u = n \cdot u_0 = 1 \; .
\label{eq:K.U}
\en
The velocity is obtained from (\ref{eq:k7}) and (\ref{eq:k8}),
\be
  u^\mu (s) \; = \; u^\mu_0 \; + \; I_1(s) \, f ^\mu _{\phantom{\mu} \nu} \,
  u^\nu_0 \; + \; \frac{1}{2} I_1^2(s) \; n^\mu  , \hbo
  I_1(s) \; = \;  a_0 \int _0 ^s d\varphi \; f(\varphi ) \; .
\label{eq:k10}
\en
Choosing initial conditions according to (\ref{eq:PULSE_dir}),
\be
  u_0^\mu = (1, \vcb{0}), \hbo n^\mu = (1, \hat{\vcb{z}}), \hbo
  \varepsilon_1^\mu = (0, \hat{\vcb{x}}) \; ,
\label{eq:k11}
\en
yields the explicit solutions for the velocity components
\be
  u^0 = 1 +  \frac{1}{2} I_1^2 \; , \quad   u^1 = -I_1 \; ,
\quad u^2 = 0 \; , \quad u^3 = \frac{1}{2} I_1^2 \; .
\label{eq:ne10}
\en
The conservation law (\ref{eq:K.U}) is seen explicitly by calculating
the longitudinal (or light-cone component) $u_\longl = u^0 - u^3 = 1$.

For a numerical solution, we evaluate (\ref{eq:n10}) iteratively: the
approximate solution in terms of the matrix $U_n(s)$ can be calculated as
long as the matrix fields $\mathbb{E}_k$ are known. Those fields, however,
depend on the particle position. Assume that we have already determined the
approximate values $u(s_i) $ for the four-velocity. We then use the trapezium
rule to calculate the position of the particle,
\be
  \bar{x}(s_i) \approx \bar{x}(s_{i-1}) \, + \, \frac{ds}{2} \Bigl( u(s_i)
+ u (s_{i-1}) \Bigr) .
\ee
Inserting the particle positions in the expression for the fields then
yields refined values for the $\mathbb{E}_k$, and an improved set $u(s_i)$
of four-velocities via  (\ref{eq:n10}). This procedure is iterated until
the particle positions and velocities at the given proper times $s_i$ do
not change within given error margins. To start the iteration we use
$u(s_i) = u_0$, for all $i = 0, \ldots , n$.

\begin{figure}[!h]
\includegraphics[width=8cm]{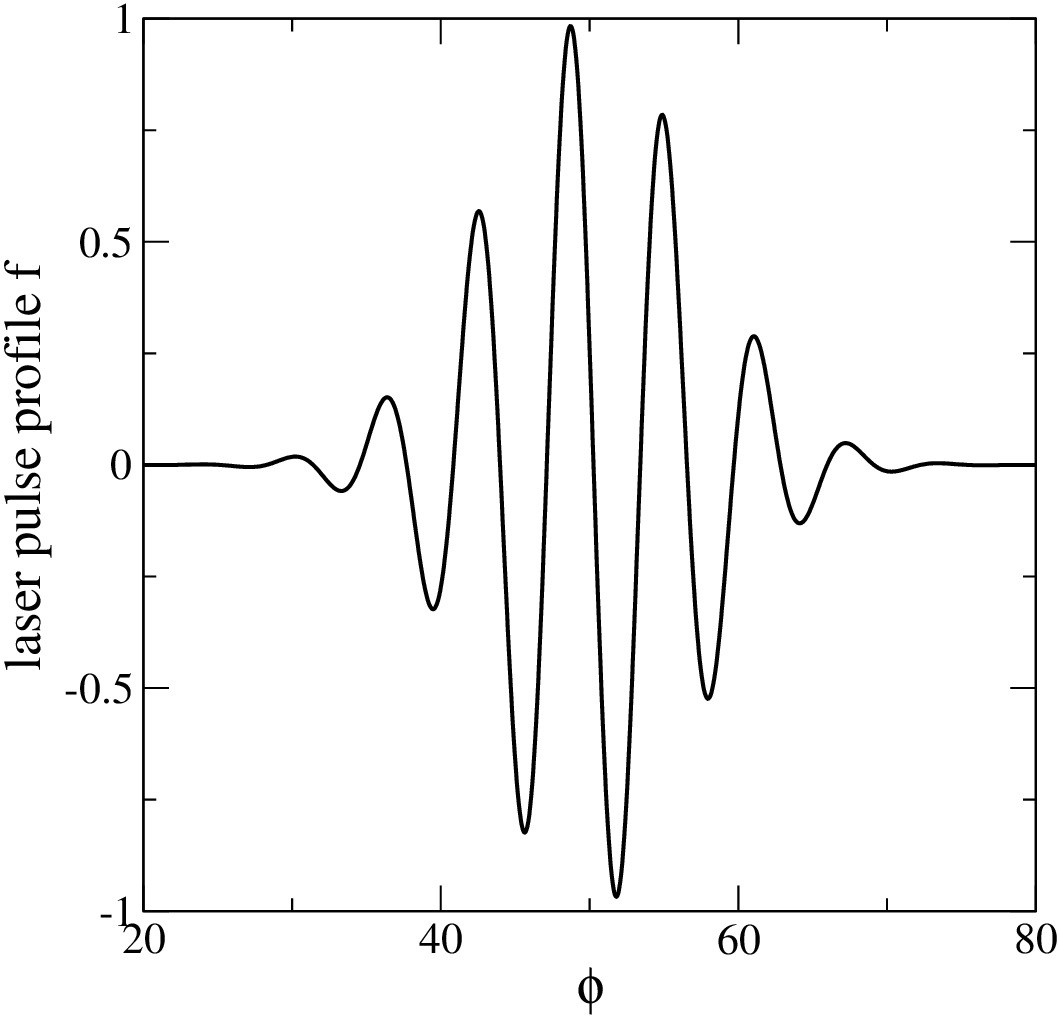}
\includegraphics[width=8cm]{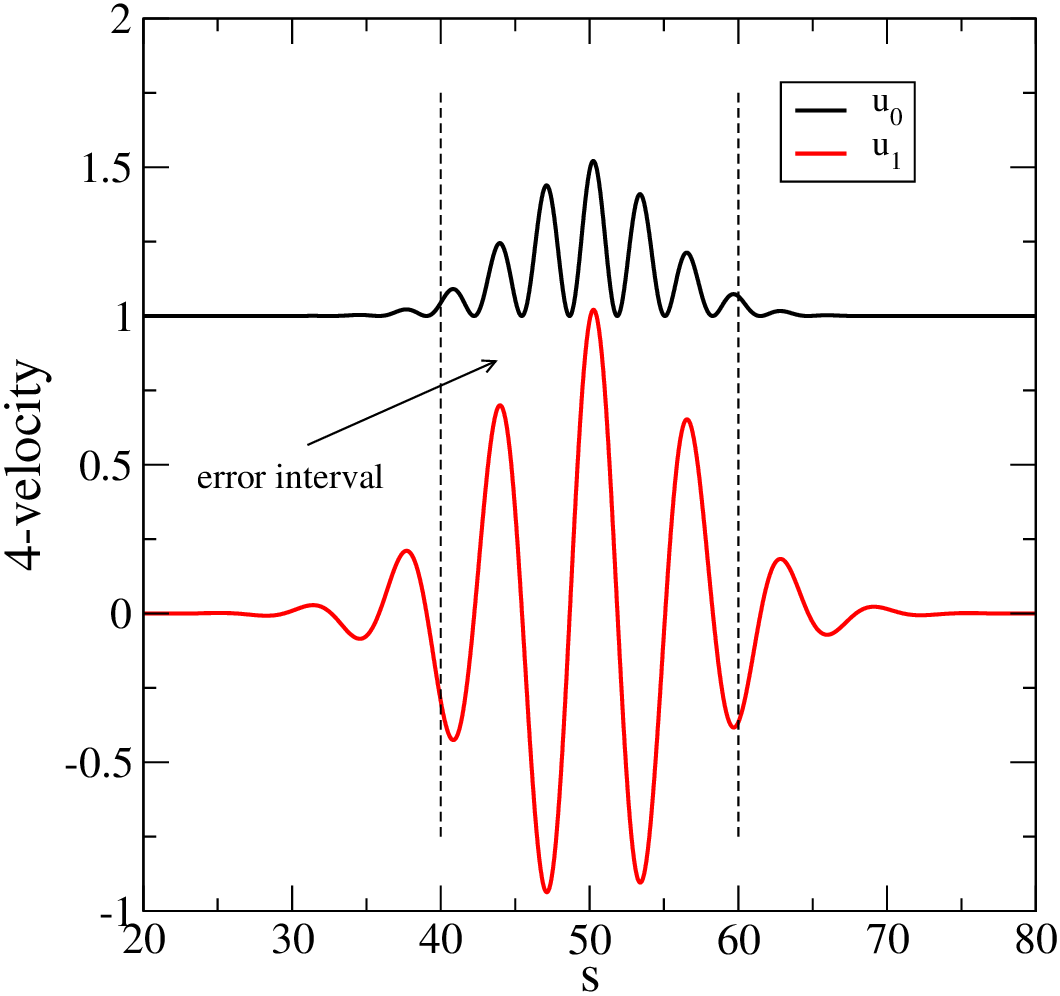}
\caption{\label{fig:1} \emph{Left:} Laser pulse function $f$ from
  (\ref{eq:PULSE}) as a function of invariant phase, $\phi$. \emph{Right:}
  Velocity components $u_0$ and $u_1$ from (\ref{eq:ne10}) as a function of
  invariant phase, $\phi = s$.}
\end{figure}

\begin{figure}[!h]
\includegraphics[width=8cm]{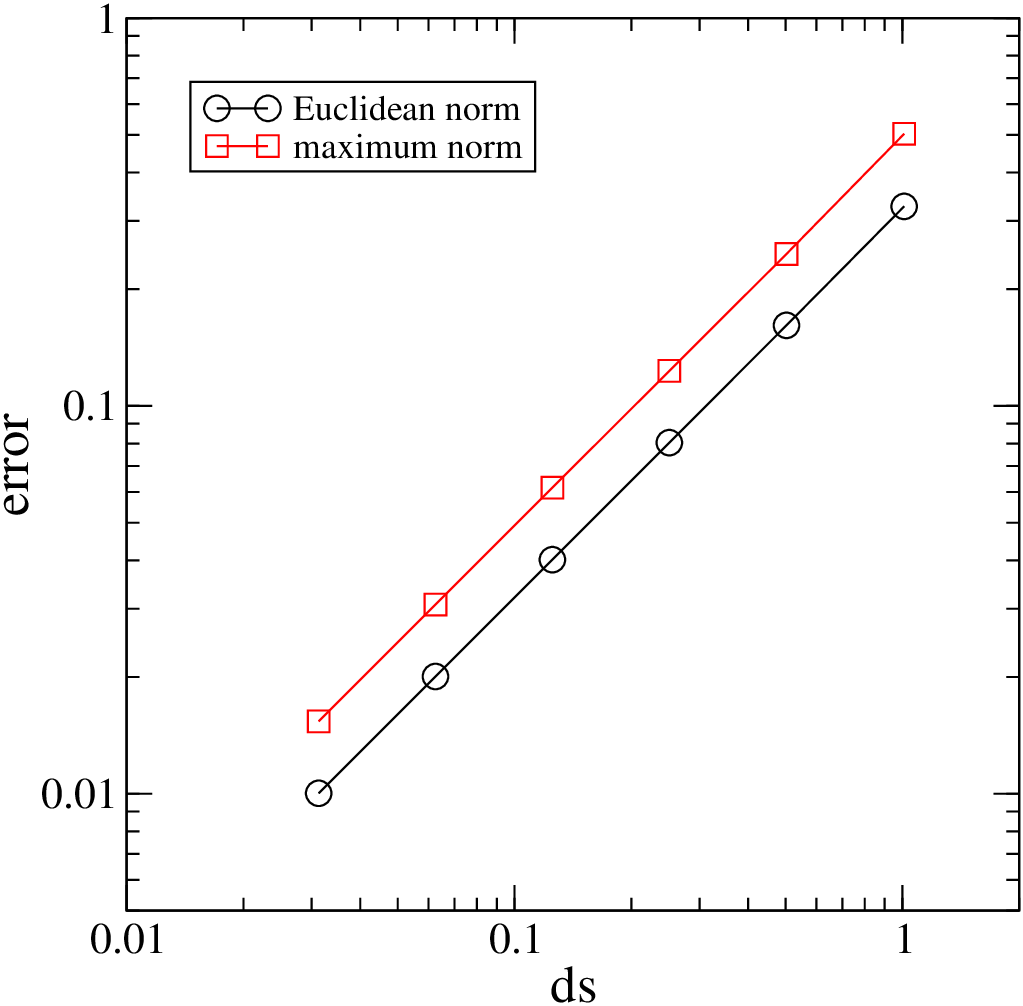}
\caption{\label{fig:2} Numerical errors (\ref{eq:NORM.E}) and
  (\ref{eq:NORM.M}) as a function of the proper time discretisation step $ds$
  for a linearly polarised laser pulse.
}
\end{figure}

\bigskip
Figure~\ref{fig:1} shows our laser pulse (left panel) and the numerical
results for velocity components $u^0$ and $u^1$ as functions of the invariant
phase $\phi$. Recall from (\ref{eq:PHI.S}) that in the absence of radiation
reaction, $\phi$ coincides with rescaled proper time, $s$. The latter has been
chosen from the range $[0,100]$ together with a step size $ds = 0.125$ and
parameter values $s_0=50$, $N =10$ and $a_0=1$. Upon comparing with the
analytic results (\ref{eq:ne10}) the numerical errors turn out to be less than
the width of the plot lines. In order to study the numerical error on a more
quantitative level, we compare the numerical solution $u(s)$ to the exact
solution, denoted $u_\mathrm{ex}(s)$. Since the numerical errors are very
small when the particle is located in the tails of the pulse, the error can be
reduced by increasing the \emph{relevant} width $\Delta s$ of the total
profile function $f(s)$. To properly characterise our pulsed plane wave and
its width $\Delta s$ we define a distribution function
\be
  \rho (s) \equiv f^2(s) \bigg/ \int_{-\infty }^\infty ds \, f^2 (s) \; ,
\ee
and the associated moments
\be
  s_k \equiv \int_{-\infty }^\infty ds \, s^k \rho(s) \; .
\ee
The width $\Delta s$ is then taken to be twice the standard deviation,
\be
  \Delta s \equiv 2 \, (s_2 - s_1^2)^{1/2} \; .
\ee
We then measure errors using both the Euclidean norm
\be \label{eq:NORM.E}
  \epsilon _\mathrm{eucl} \; = \; \frac{1}{\Delta s} \int _{s_0-\Delta s}^
  {s_0+\Delta s} \, ds \; \rho(s) \, \sum_{\mu =0}^4 \Bigl[
  u^\mu (s) -u^\mu _\mathrm{ex}(s) \Bigr]^2 \, ,
\en
and a maximum norm,
\be \label{eq:NORM.M}
  \epsilon _\mathrm{max} \; = \; \max\limits_{s, \mu}  \vert u^\mu (s)  \, - \,
  u^\mu _\mathrm{ex}(s) \vert   \; .
\en
Both errors (\ref{eq:NORM.E}) and (\ref{eq:NORM.M}) are presented in
Figure~\ref{fig:2} as a function of discretisation step size $ds$.
They are well fitted by
$$
  \epsilon _\mathrm{eucl} \approx 0.32(1) \; ds , \hbo
  \epsilon _\mathrm{max} \approx 0.50(1) \; ds
$$
and, as expected, grow linearly with $ds$.


\subsection{Pulsed Plane Wave with Radiative Reaction}
\label{subsec:RAD.NUMERICAL}

Having gained confidence in our new method it is due time to extend its
application to include radiative backreaction. As before, we assume linear
polarisation,
i.e.\ $F^{\mu\nu}(\phi) = a_0 f (\phi) f^{\mu\nu}$, cf.\ (\ref{eq:F.MU.NU}).
We then write the LL equation (\ref{eq:LL.DIMLESS2}) in terms of an
\textit{effective} field strength tensor $G^\mu _{\phantom{\mu}\nu}$,
\bea
  \dot{u}^\mu &=&  G^\mu _{\phantom{\mu} \nu} u^\nu
  \label{eq:rc1} \\
  G^\mu _{\phantom{\mu} \nu} &=& a_0 f \, f^\mu _{\phantom{\mu} \nu} + r_0 a_0
  \, u_\longl \,
  \left\{ f' \, f^\mu _{\phantom{\mu} \nu} + a_0 f^2 (n^\mu u_\nu - n_\nu  u^\mu )
  \right\} \; .
\label{eq:rc2}
\ena
Obviously, the tensor $G^\mu _{\phantom{\mu} \nu}$ also depends on the 4-velocity
of the particle, $G^\mu _{\phantom{\mu} \nu} = G^\mu _{\phantom{\mu} \nu} (\phi; u)$.
However, this does not prevent us from defining an $su(2)$ matrix
\be
  \mathbb{G} \equiv G^a \sigma_a \equiv \frac{1}{2} \left(G^{0a} +
\frac{i}{2} \epsilon^{abc}G^{bc} \right) \sigma_a \; ,
\ee
such that the LL equation can be rewritten as
\be
  \dot{U} = \mathbb{G}^\dagger U + U \mathbb{G} \; ,
\ee
in complete analogy with the $SL(2, \mathbb{C})$ Lorentz equation
(\ref{eq:SL2C.EOM}). Hence, replacing $\mathbb{E} \to \mathbb{G}$
we can again apply the iterative approach discussed in
Subsection~\ref{subsec:SL2C}. As long as $r_0$ is a small parameter the
iteration again converges rapidly.

\begin{figure}[!h]
\includegraphics[height=7cm]{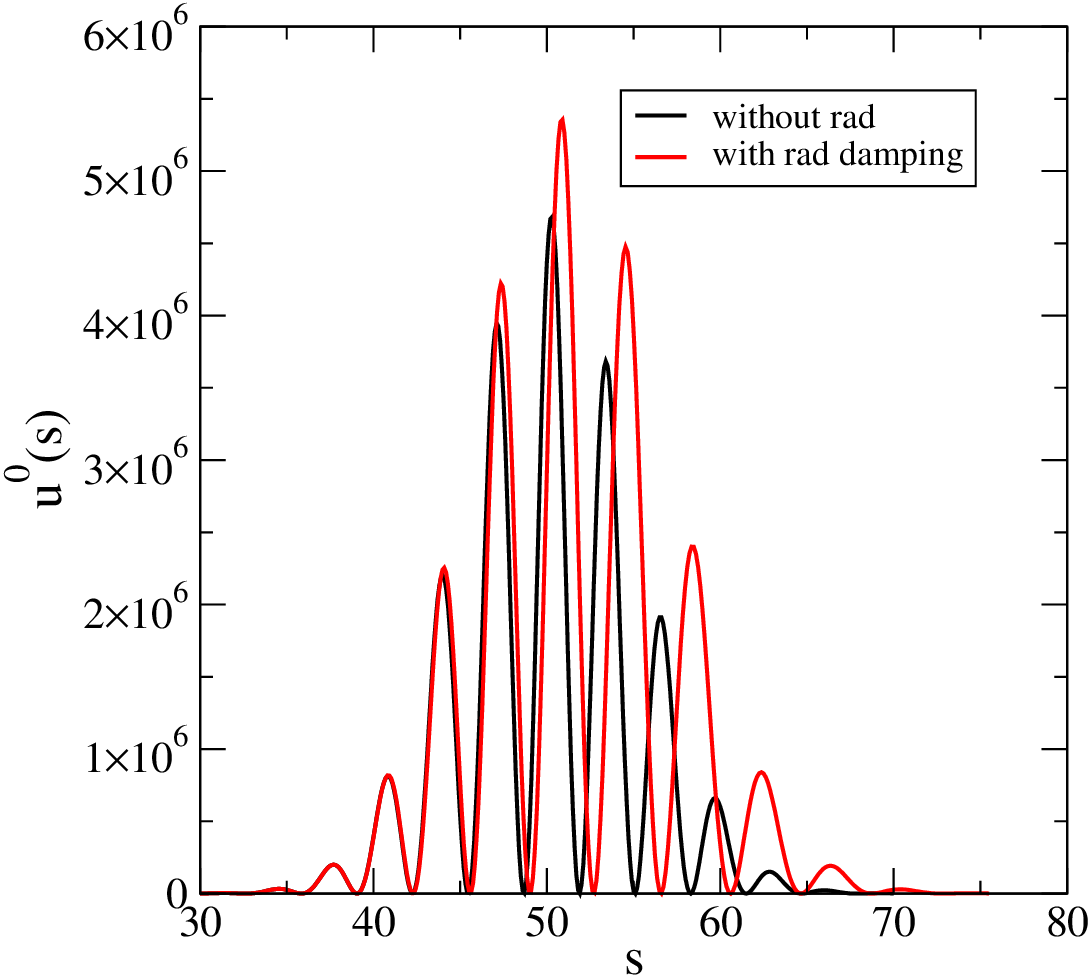}
\includegraphics[height=7cm]{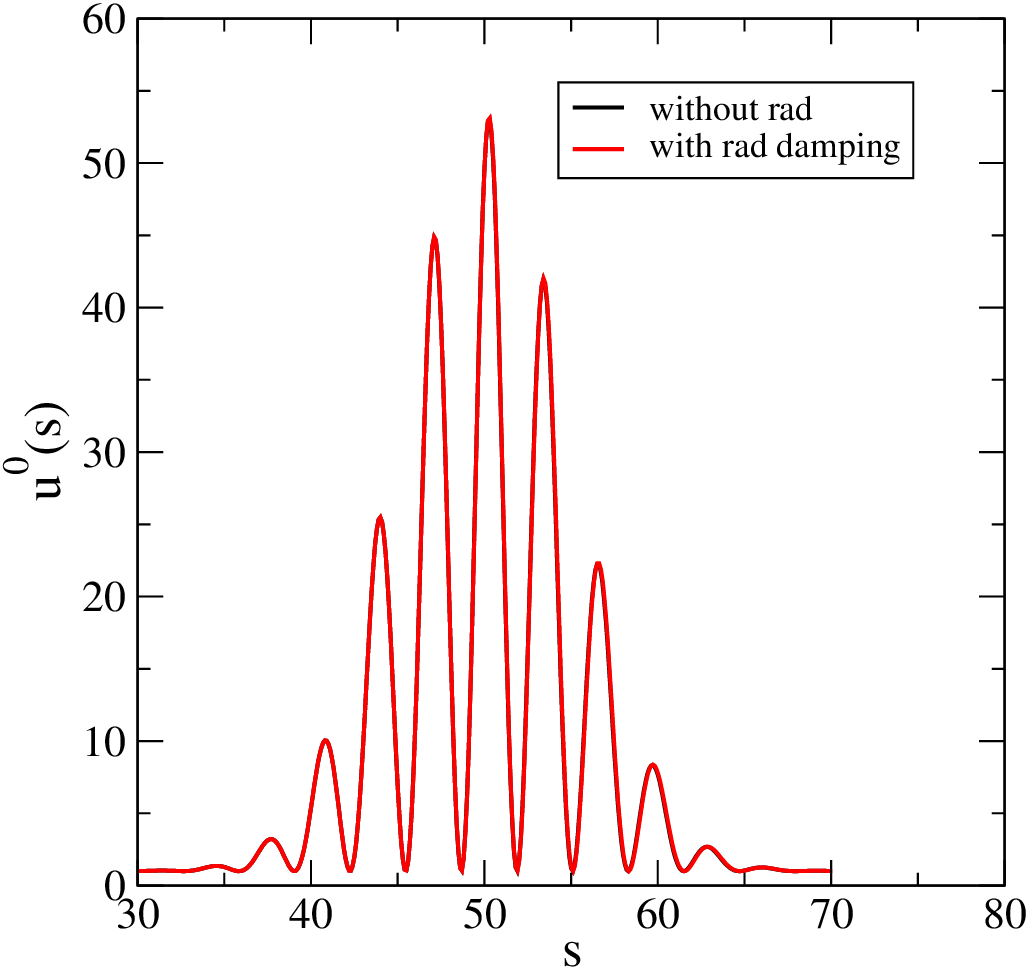}
\caption{\label{fig:RR} The $\gamma$ factor $u^0$ of the particle as a
function of the (rescaled) proper time $s$ without and with
radiative damping. \emph{Left:}
$a_0= 3 \times 10^3$ and $\nu_0 = 10^{-6}$ (optical laser). \emph{Right:} $a_0=10$ and
$\nu_0 = 10^{-3}$ (XFEL).
}
\end{figure}

In Fig.~\ref{fig:RR} we present our results for the velocity component $u^0
(s)= \gamma (s)$, the instantaneous gamma factor of the particle which
measures its instantaneous energy in units of $mc^2$. We have adopted
parameter values $a_0 = 3 \times 10^3$ and $\nu_0 = 10^{-6}$ (left panel)
as well as $a_0=10$ and $\nu_0 = 10^{-3}$ (right panel).
These roughly correspond to an optical laser of the 100 PW class envisaged for ELI \cite{ELI:2009}, and the final stage x-ray free electron laser (XFEL) at DESY,
respectively. It is clearly seen that radiative damping has a significant
effect only for \emph{optical} lasers at ultra-high intensity. Interestingly, in the lab frame where the electron is at rest initially, the radiation reaction leads to an increase of the energy amplitude. This is corroborated by the analytical solution, (\ref{eq:k7}) and (\ref{eq:k8}). The sign of the effect is consistent with the observation in \cite{Landau:1987}, Ch.~76, that the world line integral of the reaction force is the negative of the radiated total four-momentum. For a head-on collision of a charge and an infinite plane wave, the situation is different: the energy decreases in the lab frame \cite{Hadad:2010mt}.

\begin{figure}[!h]
\includegraphics[height=8cm]{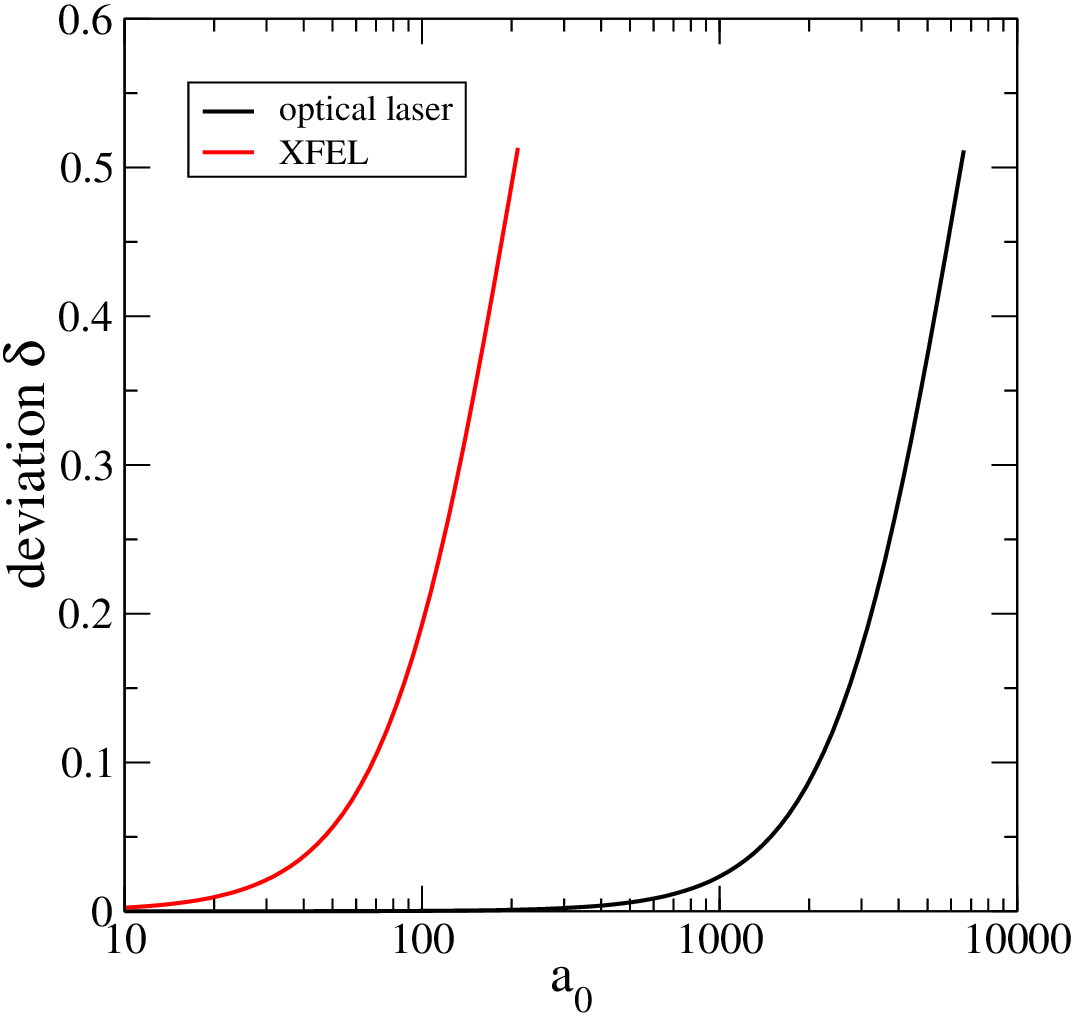}
\caption{\label{fig:4} The deviation $\delta$ from (\ref{eq:rc11}) measuring
the difference between the 4-velocities with and without radiative damping as
a function of $a_0$ for the linearly polarised laser pulse. For an optical
laser: $\nu_0 = 10^{-6}$; for an XFEL: $\nu_0 =10^{-3}$.
}
\end{figure}

In order to quantify the effect of radiation damping we view the
four-velocity as a function of both $s$ \textit{and} the fine structure
constant $\alpha = 1/137$, $u = u(s, \alpha)$. Thus, $u(s, \alpha)$
and $u(s, 0)$ represent the solutions with and without radiation
reaction (i.e.\ of LL and Lorentz equation), respectively. Denoting their
difference by $\Delta u (s) \equiv u(s, \alpha) - u(s, 0)$  we
define the maximum norm
\bea
  \delta &\equiv& \frac{1}{N} \,  \max\limits_{s, \mu} \, \left\vert \Delta
  u^\mu (s) \right\vert\; , \hbo s \, \in \, [s_0-\Delta s, s_0 + \Delta _s]
  \; ,
  \label{eq:rc11} \\
  N &\equiv& u^{\mu _\mathrm{max}} (s_\mathrm{max}, 0 ) \; ,
\eea
where $s_\mathrm{max}$ and $\mu _\mathrm{max}$ are the arguments
for which $|\Delta u^\mu (s)|$
becomes maximal. Clearly, the deviation $\delta $ is the maximum relative
difference between the velocities with and without radiative back-reaction.
It is displayed in Figure~\ref{fig:4} as a function of the invariant laser
intensity $a_0$ for the case of an optical laser ($\nu_0 = 10^{-6}$) and
the XFEL ($\nu_0 = 10^{-3}$).
The magnitude of the deviation suggests that it is controlled by the parameter $R = r_0 a_0^2$ which is $5 \times 10^{-2}$ ($5 \times 10^{-4}$) for the optical (X-ray) laser of Fig.s~\ref{fig:RR} and \ref{fig:4}.

\section{Summary and Conclusion}
\label{sec:CONCL}
In this paper we have presented a novel numerical formulation for calculating
the motion of classical charges in electromagnetic fields, with a view to
studying the behaviour of electrons in high-intensity laser beams.  Since such
dynamical systems are relativistic, one desires a formulation that is fully
covariant.  Our method employs the fact that motion in constant
electromagnetic fields proceeds along Lorentz transformation
orbits. Representing the analogue Lorentz group by \textit{space-time
  dependent} $SL(2,\mathbb{C})$ matrices we are able to numerically describe
the motion in arbitrary fields by iterative methods.  As a result, we maintain
explicit covariance and, in particular, precisely preserve the on-shell
condition, $u^2=c^2$.  We stress that the latter holds notwithstanding the
discretisation of proper time, which is required for any kind of differential
equation solver.  Conventional finite difference schemes, however,
introduce discretisation errors that violate Lorentz covariance.  Of
particular importance is the fact that our matrix formalism, by iteration, is
capable of including the radiative back-reaction on the particle motion.  To
this end we have incorporated the radiative correction terms into an effective
field strength tensor, and solved the Landau-Lifshitz equation for the test
case of a pulsed plane wave. The known analytic solution
\cite{DiPiazza:2008lm,Hadad:2010mt} is reproduced to a high accuracy. The errors scale linearly with the discretisation
step size, as one would expect for a first order method. Our results show that
radiation reaction plays an important role in an optical laser set-up at
$a_0 \sim\mathcal{O} ( 10^3)$ (while being negligible for an XFEL).

We are now in a position to study more complex field configurations such as
standing waves or more realistic models of laser beams with nontrivial
transverse intensity profiles. In particular, one may study the effects of the
laser induced mass shift \cite{Sengupta:1952,Kibble:1965zz} without having to
worry about contaminations due to discretisation errors. This requires an
appropriate (possibly numerical) definition for proper time averages in pulses
to continue the study of finite size effects on processes such as nonlinear
Thomson/Compton scattering \cite{Harvey:2009ry,Heinzl:2009nd} or laser induced
pair production \cite{Heinzl:2010vg}. There the question arises whether the
classical backreaction has a quantum counterpart \cite{DiPiazza:2010mv}. A
closely related issue is the analysis of electromagnetic photon and pair
cascades  \cite{Bell:2008zzb,Fedotov:2010ja,Sokolov:2010am,Elkina:2010up} the
details (and possibly occurrence) of which may depend sensitively on the
magnitude of radiation reaction. The discussion of all this will have to be
postponed to forthcoming publications.

\acknowledgements

The authors thank Anton Ilderton, Martin Lavelle, Mattias Marklund, David
McMullan, Hartmut Ruhl and Thomas St{\"o}hlker for fruitful
discussions. C.H. and N.I.\ acknowledge financial support through an EPSRC DTG
Scholarship. The bulk of this work has been done while C.H.\ was a research
assistant at the University of Plymouth.

\end{document}